# Solvent-Free Silsesquioxane Self-Welding for 3D Printing Multi-Refractive Index Glass Objects

**Piaoran Ye,[1]† Zhihan Hong,[1]*† Douglas A. Loy,[2,3] Rongguang Liang[1]***

[1]Wyant College of Optical Sciences, The University of Arizona, 1630 E. University Blvd, Tucson, Arizona 85721, USA

[2]Department of Chemistry&Biochemistry, The University of Arizona, 1306 E. University Blvd, Tucson, Arizona 85721-0041, USA

[3]Department of Materials Science&Engineering, The University of Arizona, 1235 E. James E. Rogers Way, Tucson, Arizona 85721-0012, USA

[*]Corresponding author: Zhihan Hong (zhihanhong@arizona.edu)

Rongguang Liang (rliang@optics.arizona.edu)

† These two authors contributed equally to this work.

## ABSTRACT

The growing interest in 3D printing of silica glass has spurred substantial research efforts. Our prior work utilizing a liquid silica resin (LSR) demonstrated high printing accuracy and resolution. However, the resin's sensitivity to moisture posed limitations, restricting the printing environment. On the other hand, polyhedral oligomeric silsesquioxane (POSS)-based materials offer excellent water stability and sinterless features. Yet, they suffer from relatively high shrinkage due to the presence of additional organic monomers. In this study, we present a polymeric silsesquioxane (PSQ) resin with reduced shrinkage, enhanced moisture stability, and the retention of sinterless features, providing a promising solution for achieving high-resolution 3D printing of glass objects. Leveraging the two-photon polymerization (2PP) method, we realized nanostructures with feature sizes below 80 nm. Moreover, we demonstrate the tunability of the refractive index by incorporating zirconium moieties into the resin, facilitating the fabrication of glass micro-optics with varying refractive indices. Importantly, the self-welding capability observed between two individual components provides a flexible approach for producing micro-optics with multiple components, each possessing distinct refractive indices. This research represents a significant advancement in the field of advanced glass manufacturing, paving the way for future applications in micro- and nano-scale glass objects.

## Introduction

Silica glass stands as one of the most valuable materials for a wide range of applications, including optics (ranging from meters to micrometers),[1, 2, 3] photonics,[4, 5] microfluidics,[6, 7] insulation[8, 9]. Its outstanding optical transparency and high thermal and chemical resistance make it highly desirable. However, the high glass transition temperature ($T_g$) of silica glass (>1000°C) poses challenges in manufacturing objects with complex 2.5-dimensional (2.5D) or 3D structures. The need for a feasible method to fabricate silica glass has spurred increasing attention on 3D printing of transparent glass in both academia and industry.

Various 3D printing techniques have been explored for manufacturing silica glass across dimensions ranging from meters to hundreds of nanometers.[10, 11] Among these techniques, photo-based printing methods,[12, 13, 14, 15] especially two-photon polymerization (2PP),[16, 17, 18, 19, 20] have consistently demonstrated superior printing resolution compared to other methods such as fused deposition manufacturing (FDM)[21] and direct ink writing (DIW).[22, 23]

One widely used approach to achieve photo-printed silica glass is through the utilization of a particle-based resin, which involves a mixture of fumed silica nano-powder, photo-curable monomers, and crosslinkers.[12] The resulting silica/polymer composite can be converted to transparent silica glass through a debinding-sintering process. This particle-based strategy offers several advantages, including easier printing of larger and thicker objects (in centimeters), non-synthetic preparation, and high stability in a standard ambient environment during printing. However, it also has several limitations. First, it necessitates high-temperature treatment (1100 - 1300°C) due to the sintering step, and its printing resolution is constrained compared to molecular resin due to the size of the particles in the resin.[20, 24] Furthermore, the difference in refractive index (RI) between the silica particles and the organic resin can introduce challenges to the printing resolution, potentially causing scattering effects,[15] which potentially restrict the applications in the high-precision optical field. Moreover, achieving the desired refractive index (RI) of glass manufactured through particle-based strategies is a nontrivial task. While powders with varying RIs can be developed, matching the RI between the powder and organic monomer (crosslinker) remains a significant challenge.

In contrast, the molecular-level resin offers an alternative strategy for 3D printing glass without employing silica particles. Several studies have reported the use of photo-based molecular level organic/inorganic hybrid resins to print inorganic glass.[13, 14, 17, 18, 19, 20] Our previous work involved the use of a solvent-free liquid silica resin (LSR) to print glass objects through 2PP, which were subsequently converted into transparent glass with optical properties nearly identical to fused silica glass at a low temperature of 600°C, although there is an additional 4% shrinkage when the glass objects were further heated to 1000°C.[18, 19] However, we recently observed that the remaining alkoxide (Si-OMe) group and silanol (Si-OH) group in LSR make it sensitive to moisture in the air, leading to gelation within a few hours when printing is conducted under humidity levels exceeding 30%. This sensitivity to moisture restricts the printing time, with higher humidity accelerating the self-gelation process. One approach to address this water sensitivity is to employ an organic/inorganic hybrid resin devoid of functional groups that are sensitive to water. J. Bauer et al. reported the use of a water-stable polyhedral oligomeric silsesquioxane (POSS) combined with a photo-reactive acrylic oligomer solvent to print glass objects with a resolution of less than 200 nm.[20] Additionally, this POSS-based resin enables full glass condensation at a mere 650°C. However, the glass objects manufactured using this POSS-based material exhibit relatively high shrinkage (~42% in the linear direction), which may limit manufacturing accuracy.[25] Moreover, the effective adjustment of the RI of the final glass has not been conclusively demonstrated in the reported study.

In this study, we present an alternative approach by utilizing a solvent-free polymeric silsesquioxane (PSQ) for the 3D printing of inorganic glass through 2PP. Our printed resin, with reduced organic content compared to previous POSS approaches, exhibits a reduced shrinkage of 32%. By optimizing the structure of the silsesquioxane with lower hydrocarbon concentrations, we further minimize the shrinkage. Additionally, we demonstrate the linear tunability of the refractive index of the final glass by introducing zirconium moieties into the resin through a simple physical mixing. The high printing resolution achieved with our PSQ is exemplified through the manufacturing of nanostructures with feature sizes below 80 nm. We successfully fabricate glass micro-optics with varying refractive indices and high accuracy. Micro-imaging systems using printed glass micro-optics were assembled and evaluated. Moreover, due to the molecular structure similarity between PSQs with different refractive, glass micro-optics containing components with different refractive indices were fabricated utilizing the self-welding process.

## Results and Discussion

### PSQ synthesis and the shrinkage during pyrolysis

To prepare a resin more stable than LSR, one approach is to replace all the tetramethoxysilane (TMOS) with silanes that only contain three alkoxy groups. This substitution ensures that all the starting silane molecules have only three functional groups each, which can be hydrolyzed instead of four. When fully hydrolyzed and condensed with trimethoxy or triethoxy groups, the resulting molecule is a well-known compound known as silsesquioxane. Among the various types of silsesquioxanes, polyoctahedral silsesquioxanes (POSS, Fig. 1a) are particularly renowned. These cage-like molecules usually consist of 8, 10, or 12 silicon atoms forming a cubic structure.[26]

The reported glass printing resin based on POSS (acrylic polyoctahedral silsesquioxane) exhibited a drawback with high shrinkage (~42%) after thermal treatment. This can be attributed to two main factors associated with the POSS-based material. Firstly, compared to our previous LSR material, the reported POSS-based material contains a higher concentration of methacrylate functional groups, resulting in a larger volume occupied by the organic component. Consequently, during the thermal treatment process, the removal of all crosslinked methacrylate chains leads to a considerable shrinkage in the material. Secondly, the highly ordered cage-like structure of POSS molecules causes them to be compactly stacked, resulting in the material having high viscosity. According to the datasheet of the acrylic polyoctahedral silsesquioxane (MA0736, Hybrid Plastics), the pure acrylic POSS without any solvent is a gel-like material with a high viscosity of 65900 mPa·s. It is worth noting that the majority of POSS compounds exist in a solid state rather than a liquid state. To conduct the printing, 9% of liquid methacrylate monomer was mixed with POSS as a reactive solvent. Although this approach allowed for 3D printing, it further increased the volume of the organic component, ultimately contributing to the final shrinkage issue.[20]

To minimize shrinkage, a key approach is to reduce the viscosity of silsesquioxane without the need for additional monomers or solvents. In this regard, the utilization of PSQ proves to be an ideal solution. Unlike traditional POSS, polymeric silsesquioxane (PSQ)[27] is a mixture comprising a range of random ring-structured silsesquioxanes with varying molecular weights (Fig. 1b). These flexible rings, along with organic functional groups attached to them, act as spacers, resulting in a mixture with lower viscosity compared to POSS. For the synthesis of PSQ, 3-methacryloxypropyltriethoxysilane (MPTES) was selected as the primary monomer. The synthesis process employs an $F^-$ catalyzed mechanism, enabling the formation of a fully hydrolyzed and condensed product within a short span of 2 hours. After removing all solvents used during the synthesis, the final product is a viscous oil, which presents no difficulties in dissolving photo-initiators without the need for any additional solvents. This advancement opens up possibilities for reducing shrinkage in the resulting material.

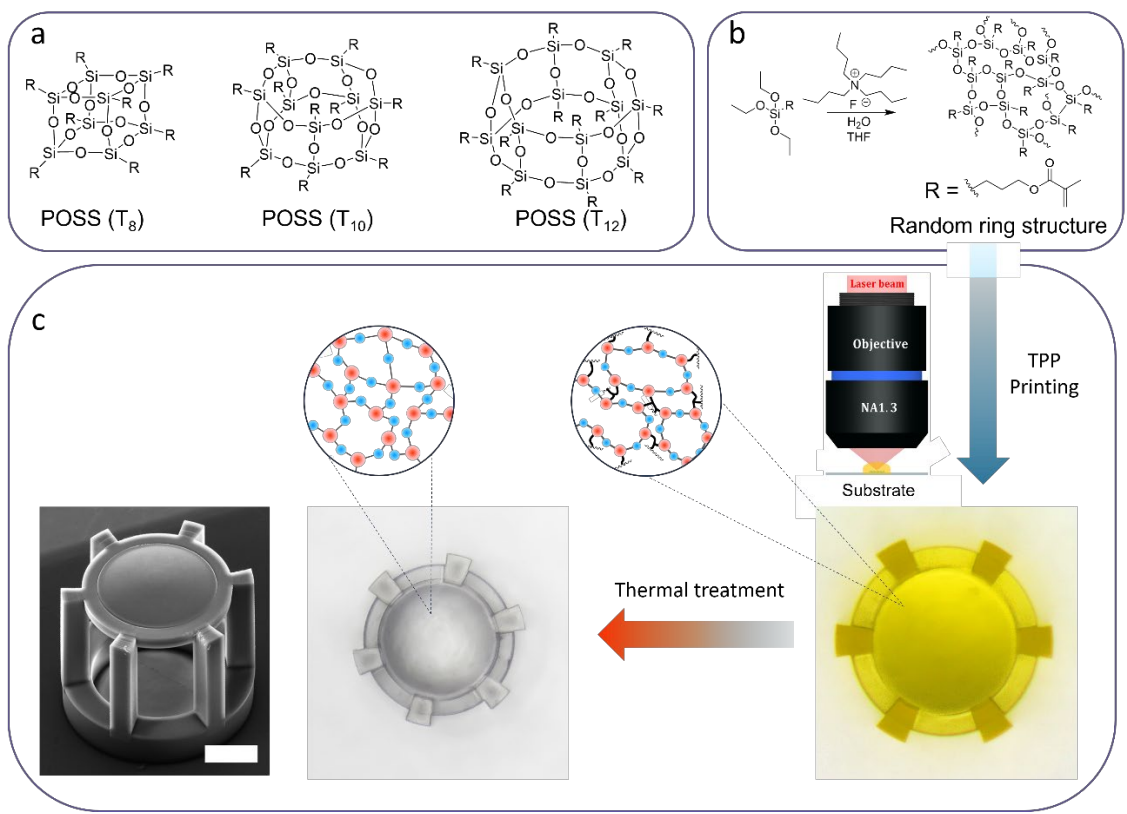

Figure 1. (a) Typical structures of POSS molecules with different cage sizes. (b) Scheme of synthesizing PSQ through a $F^-$ catalyze route. (c) Scheme of a basic procedure to print and get inorganic glass objects after thermal treatment. The SEM image shows a printed glass single lens. Scale bar: 100μm.

Before measuring the shrinkage of our PSQ, we conducted TGA (Thermogravimetric Analysis) to gain a better understanding of the thermal behavior of UV-cured silsesquioxane and to optimize the thermal treatment procedure. We observed that to achieve complete pyrolysis of the cured PSQ, a temperature higher than 600 ℃ (approximately 650 ℃) is required. This temperature is similar to the one reported for printed POSS material[20] but slightly higher than that of our previously reported LSR material, which only required 600 ℃ for complete pyrolysis. The reason for this difference lies in the higher ratio of methacrylate groups and the higher crosslinking density present in both the PSQ and POSS during UV curing. In the case of photocrosslinked acrylates, a higher crosslinking density typically contributes to enhanced thermal stability, leading to higher degradation temperatures.[28]

To assess the shrinkage of the PSQ prepared using MPTES, we fabricated a lens structure via 3D printing and subjected it to a thermal treatment at 650 ℃. Both Raman (Fig. S1b) and FTIR (Fig. S1c) spectroscopy analyses revealed that the sample treated at 650 ℃ exhibited a chemical structure similar to fused silica.

The dimensional change observed indicated that approximately 33% of shrinkage occurred (Fig. S2c), which is nearly 10% less compared to the shrinkage observed in the case of POSS-based material. This notable reduction in shrinkage can be primarily attributed to the absence of organic monomers in our PSQ formulation.

To further reduce shrinkage, we adopted a second strategy, which involved replacing the long methacryloxypropyl group with a shorter methacryloxymethyl group. In a previous study,[18] we demonstrated that shorter side chains in LSR resulted in lower shrinkage after thermal treatment, primarily due to the reduced organic volume. Building upon this knowledge, we utilized 3-methacryloxymethyltrimethoxysilane (MMTS) as the starting monomer to synthesize PSQ using the same procedure as before (Fig. S2a).

Thanks to the shorter organic side chain, this material exhibited an even lower shrinkage, measuring only 28.5%. This result aligns well with the TGA findings, as the PSQ-MMTS demonstrated 46.2% weight remaining after being subjected to 650 ℃, which is 4.3% higher than the TGA results for PSQ-MPTES and 11.5% higher than the results for POSS-based material (Figure S2b). One thing that needs to be mentioned is that the PSQ-MMTS tends to self-polymerize when it is placed at room temperature, which is not an obvious phenomenon for the PSQ-MPTES. The addition of the inibitor into the final resin helps suppress this self-polymerization process.

The synthesized PSQ exhibits high stability to water when compared to our previously published LSR. The absence of methoxy or hydroxy groups in the final PSQ renders the material highly stable, allowing PSQ-MPTES to be stored at room temperature for at least one year without gelation. This stability makes the material inert to moisture in the air, enabling printing under various humidity conditions without the risk of self-gelation. In a typical printing scenario, no observable changes were noted in the material even after it was exposed to an ambient environment for a duration of 5 days.

**Reaching nano-scale printing resolution**

To demonstrate the printing resolution of our PSQ resin in 2PP 3D printing, we printed a series of grid and grating structures with varying feature sizes. Figure 2a – 2c showcase a printed micro-cubic grid structure supported by a framework, featuring sharp edges and well-defined details. Figures 2d-2f exhibit printed grating structures ranging from micrometers down to 75 nm, achieving some of the highest printing resolutions observed in 2PP printing. Of particular interest is Figure 2e, where the gap between each grating line measures approximately 430 nm, and no merging of adjacent lines was observed, underscoring the high spatial printing resolution attained.

Moreover, in Figure 2g, we demonstrate precise control over nanometer-sized grating structures with the desired width. An array of gratings with varying line widths, from thin (left) to thick (right), was printed, and upon exposure to white light, different colors were observed due to the diffraction caused by varying line widths. Meanwhile, it is essential to mention that the light source we used was not parallel for all the grating structures, resulting in slightly different input angles for each grating cube, which, in turn, influenced the colors observed in the diffraction pattern.

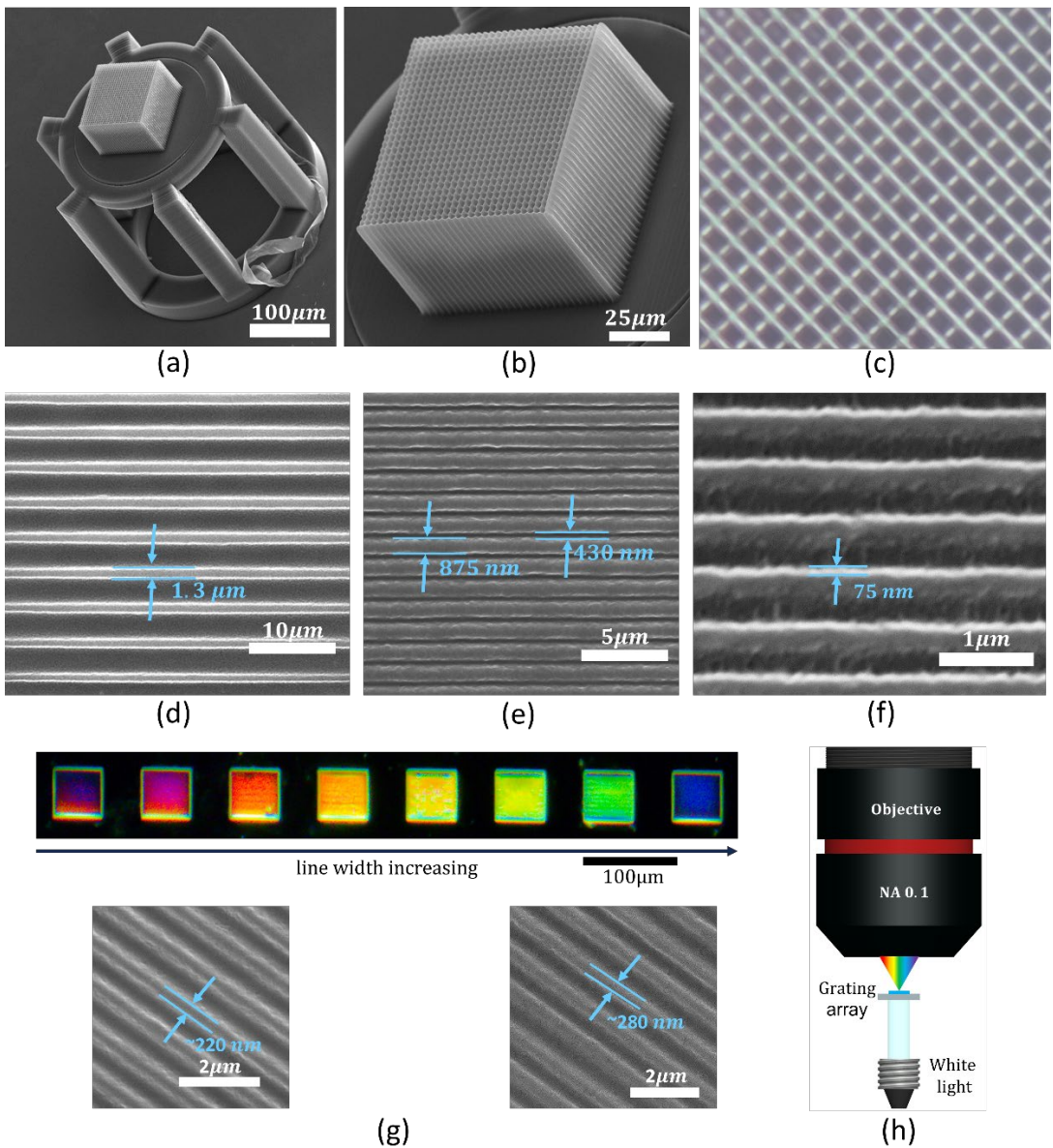

Figure 2. (a)-(b) SEM images of a micro-grid structure printed using PSQ. (c) microscopic image of the same grid. (d)-(f) 3D printed structures with featured size and pitch size from micrometer to nanometer. (g) microscopic image of 3D printed grating array with different featured size. The color was caused by diffraction of the white light. The image was captured under the dark field. (h) The instrumental scheme of the diffraction experiment.

**Change the refractive index of PSQ and final glass**

The remarkable stability of our PSQ resin opens up the possibility of tuning the refractive index (RI) of the final glass by introducing metal alkoxide molecules. In previous literature, it has been reported that the refractive index can be adjusted by chemically bonding zirconium alkoxide into a methacrylate-functionalized silsesquioxane network through a sol-gel process, resulting in a 3D-printable ink capable of producing prints with tunable RI.[17, 29] However, the material obtained using this method requires the presence of solvents to maintain a liquid state and transforms into a solid gel once all the solvent is removed. This feature necessitates a specialized coating and heating procedure for sample preparation, as well as a distinct printing setup, as the oil-immersed objective cannot be used with the solid film of the material.[30] Moreover, writing directly in a solid film, rather than in a liquid, could potentially impact printing resolution and complicate post-printing development, making it more time-consuming.

To address these challenges and develop a solvent-free 3D-printable liquid resin with low viscosity, while also being resistant to moisture, we opted to stabilize the zirconium alkoxides by using methacrylic acid. Zirconium alkoxides are highly sensitive to water and tend to undergo hydrolysis upon contact. By binding carboxylic acids to the Zr core, the stability of zirconium alkoxides can be significantly improved.[29, 31, 32] During this binding process, propyl methacrylate and 1-propanol are generated as byproducts (for detailed discussion, refer to the supporting information). The resulting Zr complex was then physically dissolved into the synthesized PSQ. Subsequently, we removed propanol under vacuum, while propyl methacrylate was retained due to its high boiling point (see disscussion in supporting information). A basic scheme depicting the preparation of the RI tunable resin using the Zr complex is shown in Figure 4a.

Due to the absence of chemical bonds between Zr and silsesquioxane, this material can maintain its liquid status. It should be noted that the concentration of zirconium moiety was kept below 10 mol%. Higher concentrations were found to lead to two major issues: an increase in ink viscosity and heightened sensitivity to moisture, making the ink more prone to gel formation in the air. Additionally, through our experimentation, we observed that PSQ-MPTES (<3000 mPa-s) proved to be a more suitable base material compared to PSQ-MMTS (>16000 mPa-s), mainly due to its lower viscosity (Fig. S3). As a result, we exclusively mixed the Zr complex with PSQ-MPTES to achieve the desired refractive index tuning for the final glass.

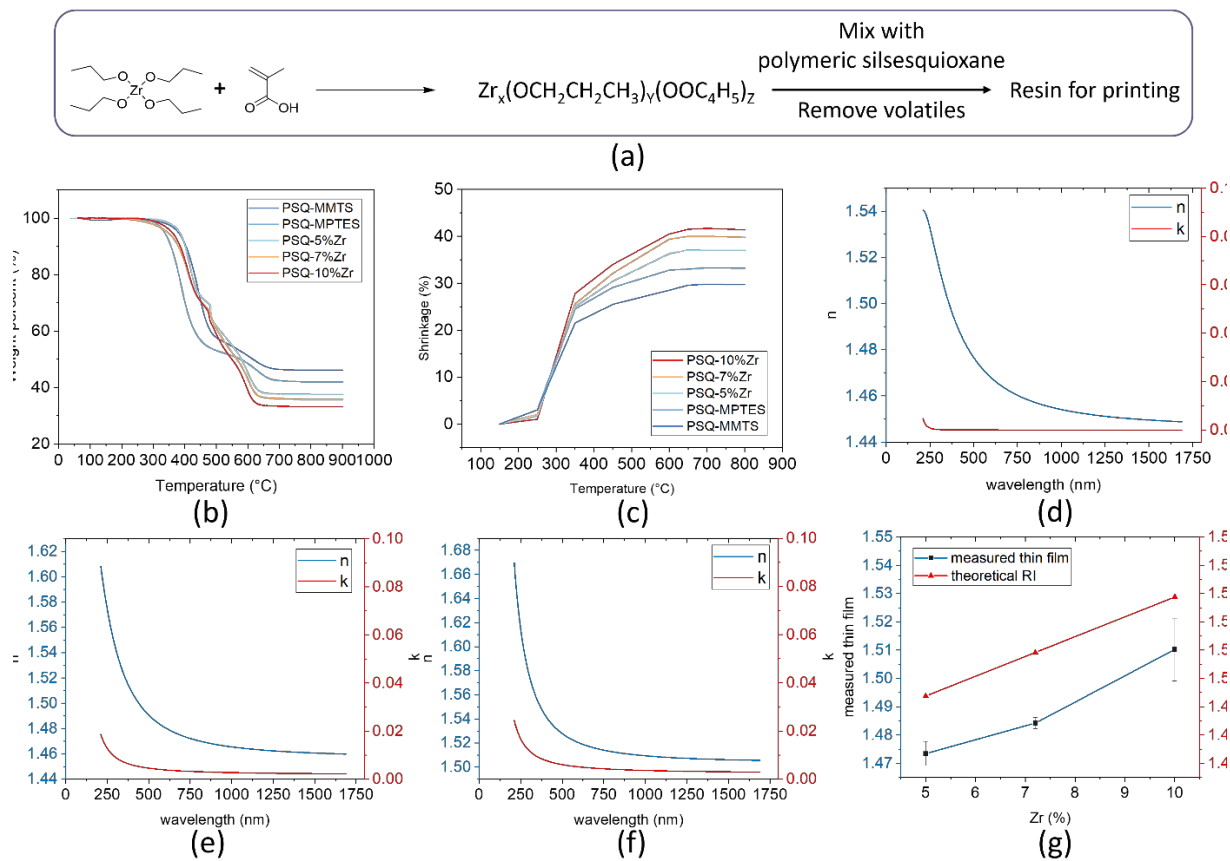

Figure 3. (a) Scheme of preparing zirconium complex stabilized by MAA and preparing the resin for glass with different refractive index. (b) TGA results of UV-cured samples with different zirconium concentrations. (c) The measured shrinkage of samples at different temperatures during the thermal treatment. (d)-(f) The refractive index, n, and extinction coefficient, k, of samples with different zirconium concentrations. (g) The refractive index of samples with different zirconium concentration at 580 nm and the corresponding theoretical RI of each sample.

The introduction of the Zr complex into the printing material brings the organic component methacrylic acid (MAA), contributing to the organic volume and increasing the shrinkage of the final glass. TGA measurements revealed that as the Zr concentration increased from 0% to 5%, 7.2%, and finally 10%, the remaining weight after thermal treatment decreased from 41.9% to 37.6%, 35.7%, and 33.3%, respectively (Fig. 3b). The samples containing the Zr complex exhibited a mass decrease at around 475℃ due to the decomposition of poly(methacrylic acid).[33] The lower remaining weight indicates a reduced weight percentage of the final glass, leading to higher shrinkage after thermal treatment. As depicted in Figure 3c, the gradual increase in Zr concentration up to 10% resulted in a final shrinkage of 41.5% at

650℃, which is approximately 8% higher than the pure PSQ-MPTES with about 33% shrinkage. In addition, the Raman spectrum of a printed sample after thermal treatment (Fig. S4) shows that there is no typical $ZrO_2$ crystal structure indicating the Zr is homogeneous dispersed before and after thermal treatment.

To measure the RI of final glass with different Zr concentration, we prepared thin films using PSQs on quartz, did the thermal treatment, and did the measurement using ellipsometer. Theoretically, the refractive index of a multi-component material can be roughly calculated using the equation showing below:

$$n_{theoretical} = \Sigma(n_i \times c_i) \qquad (1)$$

Where the $n_{theoretical}$ refers to the theoretical refractive index, $n_i$ refers to the refractive index of component i, and $c_i$ refers to the mole concentration of component i in the whole material.

Using equation 1, the theoretical $n_{580nm}$ for the final glass was calculated as 1.494, 1.509, and 1.529 when the zirconium concentration increased from 0% to 5 mol%, 7.2 mol%, and 10 mol%, respectively. Figures 3d-3f present the measured $n_{580nm}$ of the thermally treated thin films with 5 mol%, 7.2 mol%, and 10 mol% of Zr, which were determined as 1.473, 1.484, and 1.510, respectively. Although all measured values were slightly lower than the theoretical RI, they were all higher than the RI of silica at 580nm (1.459), confirming that the addition of Zr effectively increased the refractive index as intended. The lower RI compared to the theoretical value can be attributed to the films being coated and fixed on a substrate. When a free-standing object undergoes thermal treatment, shrinkage occurs in all directions due to the elimination of the organic component, even if the temperature is below the glass transition temperature ($T_g$) or sintering temperature ($T_s$). However, when the material is in the form of a thin film fixed on a substrate, the horizontal shrinkage is restricted unless cracks form, and only vertical shrinkage can occur. This also explains why the thickness of the thin film changed more compared to the shrinkage of printed free-standing objects before (1500-1700 nm) and after (400 to 600 nm) thermal treatment. Moreover, the treatment temperature in our experiments was lower than $T_s$ of most samples with different Zr concentrations, except for the sample with 10 mol% Zr.[34] Therefore, even though vertical shrinkage still occurred to a certain extent, the overall density of the film was lower compared to the free-standing sample, resulting in a lower RI. This phenomenon of low RI is common in sol-gel prepared thin films on substrates.[35, 36] Additionally, it should be noted that the refractive index of free-standing glass optics should be higher than that of thin films.

As the Zr concentration increased, the k value of each sample also showed an increment, particularly in the UV region (<400 nm). At 5% Zr concentration, there was almost negligible absorption for wavelengths above 300 nm. With a gradual increase in Zr concentrations to 7.2% and 10%, a slight increase in absorption of visible light was observed, but the values remained low, mostly below 0.005. However, the absorption of UV light exhibited a significant increase, reaching a value as high as 0.024 (for 10% Zr at 210 nm). This pronounced increase in absorption in the UV region is primarily attributed to the absorption of the Zr-O structure, as silica thin films generally exhibit minimal absorption from approximately 300 nm up to the visible light range.[37, 38]

**3D printing of optics to build imaging system**

We use a 780nm femtosecond laser printing system to 3D print optics needed for assembling imaging systems. To demonstrate the high printing accuracy of our system, we printed a flat element as well as a aspherical lens structure. The surface quality of this two objects were measured after thermal treatment at 650 ℃. The low surface roughness (< 4nm, Fig S5a) and the low peak to valley deviation (± 135 nm, Fig. S5b) indicate a well-controlled surafce quality can be reached, which is essential for imaging applications.

To build the imaging systems, we printed two types of imgaing optics. One is the micro-objective structure with 3 lens structures as shown in Figure S6a. The imaging resolution of this objective itself was demonstrated using a set up as described in our previous paper.[19] Notably, the discernment of the initial three rows of elements within group 9 of the 1951 USAF target (Fig. S6b) underscores the attainment of sub-micrometer imaging resolution. This level of resolution facilitates the discernment of intricate details in human mammary gland samples (Fig. S6c). A comprehensive illustration of the imaging performance of the printed objective across various groups within the 1951 USAF target is presented in Figure S7.

Such a multi-lens design is essential and useful for medical applications. Therefore, we furhter printed a doublet glass microscope to demonstrate the potential of a minimally invasive imaging system for the detection of internal body diseases.[39] This innovation aims to pave the way for micro-detection systems tailored for organs with smaller lumens.[40, 41, 42, 43] The system's versatility extends to accommodating supplementary modalities, including precise tissue sampling, laser-based therapy, and targeted drug administration. Significantly, the preference for silica optics over conventional polymers or plastics was steered by its discernible advantages in terms of biocompatibility and superior transmission of ultraviolet/blue light. The experimental setup incorporated the printed glass objective, juxtaposed with a 10,000-element fiber bundle (PN: FIGH-10-350S, Fujikura) against the glass microscope (Fig 4a). The distal end of this fiber bundle was strategically positioned at the working distance (WD) of a microscope system equipped with a 10X objective. As evidenced in Figure 4d, the printed micro-objective resolved element 4 in Group 7. Figure 4b and 4c present images captured from a histological section of a fallopian tube and an injected rabbit lung, showing the potential for medical purposes.

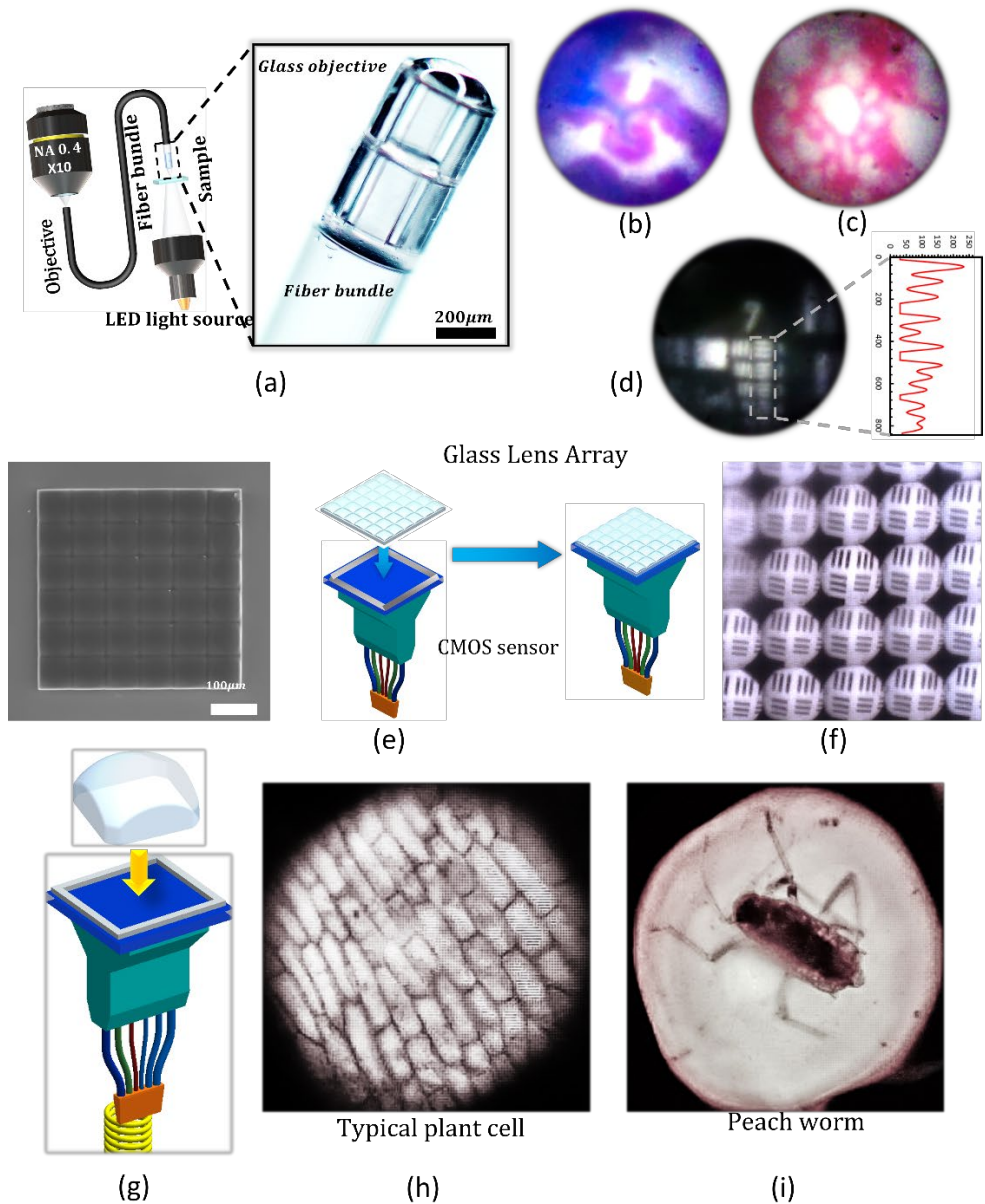

Figure 4. (a) The schematic diagram and microscopic image of the experimental setup for fiber-bundle imaging experiments. (b) The image of the histological slide of human fallopian tissue captured by the system shown in (a). (c) The image of the slide of injected rabbit lung sample. (d) The image of 1951 USAF target 7th group elements. Note that the honey cone pattern caused by fiber bundle has been removed by imaging processing using software. (e) The SEM image of the manufactured 6X6 glass micro lens array, and the scheme of the lens array on the CMOS sensor with a pixel size of 1.75 μm X 1.75 μm. (f) The image captured by the setup in (e) of a resolution target. (g) The scheme of setup fabricated using a singular lens.

(h)-(i) The images captured by the single lens mounted on CMOS sensor of bio-samples (plant cell & peach worm).

Besides the glass objective for the endoscope type application, the 2PP printed micro-glass optics also provide good opportunity for fabricating ultra-compact imaging system, such as the chip-on-tip design. Unlike the conventional endoscope design which use a fiber bundle or waveguide to transfer light signal from lens to the camera, the chip-on-tip design uses a small (submillimeter) image sensor located at the endoscope tip equipped with the lens element, making the whole system more compact than conventional endoscope system. As depicted in Figure 4e & 4g, the glass optics—whether a lens array or a singular lens—can be mounted atop the COMedia CMOS sensor, which boasts a resolution of 384x384 pixels. This chip is characterized by a pixel pitch measuring 1.75 µm. To fully utilize the sensor area of the mounted lens, each lens was architecturally designed with a square-shaped entrance pupil, ensuring comprehensive coverage of the sensor. Figure 4f presents an image of the 1951 USAF target (group 1, element 6th) as captured by a 6x6 glass lens array. It is noteworthy that the diameter of the lens array surpasses the dimensions of the available sensor area. Figures 4h and 4i display images rendered by the singlet glass lens, emphasizing their distinct features.

**3D printing of multi-refractive index micro-optics through self-welding of PSQs**

One challenge faced by glass micro-optics lies in the intricate process of assembling two or more objects together. This challenge becomes particularly pronounced when dealing with glass objects below 500 micrometers in dimension. Conventional adhesive methods, although effective in many scenarios, present a potential risk of compromising the integrity of optical components of such small scale. While there have been reports of laser welding glass without the use of additives,[44] the precise alignment and welding of micro-sized glass optics at desired positions remains a nontrivial endeavor.

One notable advantage of our printable PSQs is their inherent self-welding capability. This characteristic imparts high flexibility for constructing objects containing multiple components with varying refractive indices. When two discrete printed PSQ objects (PSQ-MPTES and PSQ-5%Zr) were positioned closely during thermal treatment, an automatic welding process occurred, without the need for sintering, melting procedures, or additional welding agents. This phenomenon, which we term "self-welding behavior" of PSQ, culminated during the pyrolysis at 650 ℃.

As illustrated in Figure 5a, the molecular structures of component 1 and 2 are similar, with the exception of the Zr moiety in component 1. Although they are not chemically bonded, the smooth surface facilitates close contact, enabling the formation of Si-O-Si bonds that effectively weld the two components together.

To assess the welding strength, we conducted tests using microfibers to manipulate and stretch the welded components, confirming their secure fusion.

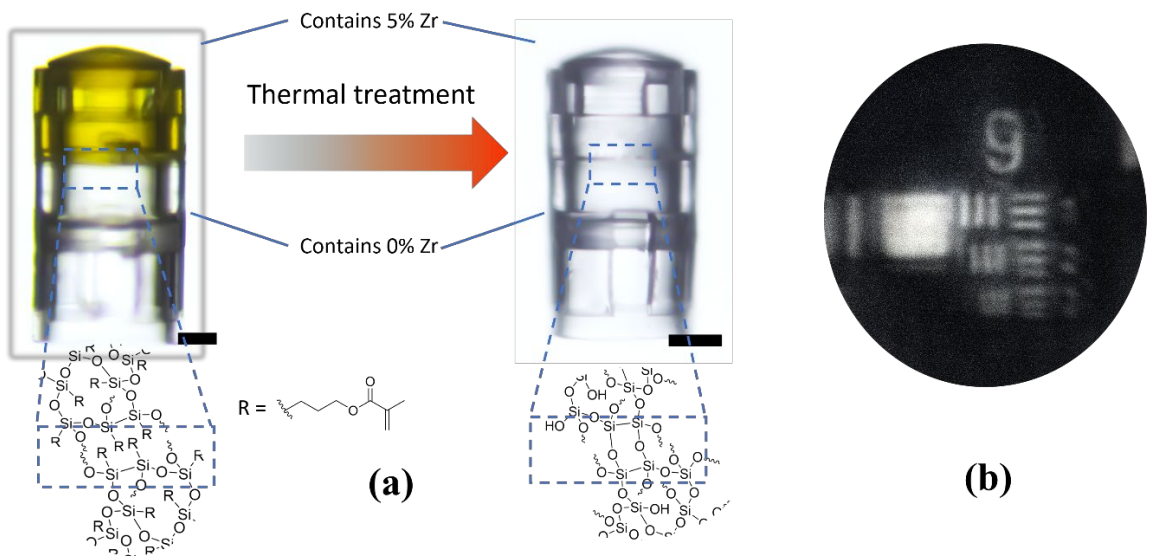

Figure 5. (a) The scheme of the self-welding process. Zr moieties were not presented in the molecular structure. The PSQ-5%Zr shows a much more yellow color compared to the PSQ without Zr, which is due to the interact between BEBP and carboxylic acid. During pyrolysis, Si-O-Si bonds were formed to weld two individual parts. (b) The image of USAF target captured with the micro-optics shown in 5a.

It is worth noting that, despite PSQ-5%Zr displaying approximately 5% greater shrinkage compared to PSQ-MPTES, we did not observe significant disparities in the outer frame. This observation may be attributed to the welding behavior, which seemingly constrained the shrinkage of the PSQ-5%Zr sample, particularly since PSQ-5%Zr exhibits similar shrinkage to PSQ-MPTES at temperatures below 400 ℃. As demonstrated in Figure 4b, the fabricated micro-optics exhibit good imaging quality, affirming the alignment's suitability for imaging applications. Nonetheless, when combining PSQs with higher Zr concentrations alongside PSQ-MPTES or PSQ-MMTS, the potential for a more pronounced mismatch exists, which could lead to alignment challenges. This issue can be addressed by introducing redundancy to components with higher shrinkage and incorporating designed position-limiting structures within the contact region. These strategic measures will serve to compensate for the higher shrinkage, thus ensuring precise alignment.

## Conclusions

In this paper, we have developed a strategy to print inorganic glass utilizing PSQ. The use of random ring structure in silsesquioxane enables 2PP 3D printing without the need for additional solvents or reactive monomers, resulting in reduced shrinkage after thermal treatment compared to POSS-based materials.

Moreover, the absence of reactive methoxy or silanol groups enhances the stability of PSQ against moisture, making it more suitable for certain applications. The similar molecular structures of PSQs with different Zr concentrations and refractive index allows self-welding of individually printed components, which further enhances the ability of this PSQ system for a wider range of applications.

The experimental results demonstrate that the silsesquioxane produced by MPTES exhibits a final shrinkage of 32.8%, which is further lowered to only 28.5% with the MMTS variant. Additionally, high printing resolutions of approximately 75nm featured size and a spatial resolution of at least 430nm (distance between two printed line structure) can be achieved by 2PP. The ability to tune the RI of the final glass through the introduction of zirconium-MAA complex opens up possibilities for a variety of optical applications. In this paper, introducing 10mol% Zr brings RI up to 1.510. The fabricated micro-optics with low surface roughness (<4 nm) exhibit excellent imaging performance, achieving high imaging resolution (sub-micrometer) and proving their suitability for real micro-sized bio-samples.

Further research efforts can be directed towards increasing the refractive index to meet the demands of various optical applications. One potential approach could involve introducing higher concentrations of zirconium into the silsesquioxane, although this may require addressing challenges related to viscosity and sensitivity to water. Considering the reported stability of acetic acid-stabilized zirconium,[31] using acetic acid might be a promising avenue to achieve higher zirconium concentrations and increase the refractive index, albeit potentially leading to higher shrinkage and necessitating monitoring for phase separation during printing and thermal treatment.

In summary, this research work represents an advancement in the preparation of solvent-free resins for 3D printing inorganic glass structures, offering valuable contributions to the fabrication of micro- or nano-sized glass objects. By addressing the limitations and exploring potential future directions, this study paves the way for the continued exploration and application of PSQ in the field of advanced glass manufacturing.

## Materials and Methods

### Materials

Methacryloxymethyltrimethoxysilane (MMTS) and 3-methacryloxypropyltriethoxysilane (MPTES) were purchased from Gelest. 4,4'-Bis(diethylamino)benzophenone (BEBP), monomethyl ether hydroquinone (MEHQ), zirconium(IV) propoxide (70 wt. % in 1-propanol), and Propylene glycol monomethyl ether acetate (PGMEA) was purchased from Sigma-Aldrich. All other chemicals were used as received unless a specific statement.

### Synthesis of polymeric silsesquioxane (PSQ) using MPTES

MPTES (5g, 0.017mol) was mixed with tetrahydrofuran (6g) in a 50 mL round bottom flask. Deionized water (0.92g, 0.051mol) was added to this solution. Under magnetic stirring, 0.17mL Tetrabutylammonium fluoride solution (1M in THF) was added dropwise. This solution was allowed to be mixed at room temperature for 2h. After that, the solution was poured into 30 mL water and extracted using 8mL of dichloromethane twice. The combined extracted solution was dried with brine and magnesium sulfide. After that, 15mg of 4,4'-Bis(diethylamino)benzophenone (BEBP) was added, and the solution was concentrated to around 5 mL. After it was filtered using a 0.02 μm Anotop filter and the remaining dichloromethane was removed under vacuum leaving a viscous yellow oil. $^1$H NMR (500 MHz, CDCl$_3$) δ = 6.10-5.96 (m, 1H), 5.53-5.45(m, 1H), 4.20-4.01(m, 2H), 1.92-1.81(m, 3H), 1.80-1.65(m, 2H), 0.75-0.60(m, 2H); $^{13}$C NMR (500 MHz, CDCl$_3$) δ = 167.3, 135.9, 124.8, 65.8, 22.1, 17.5, 7.4.

**Synthesis of PSQ using MMTS**

The procedure is the same as the procedure to synthesize the PSQ using MPTES while the MPTES was replaced with the same mole of MMTS. 5mg of MEHQ was added as inhibitor to the final resin. The synthesized PSQ-MMTS was placed in fridge for storage.

**Synthesis of PSQ with different refractive index**

To tune the refractive index of final glass, zirconium propoxide was selected as the source of zirconium to increase the refractive index. To prepare the PSQ with different zirconium concentrations, the PSQ was firstly synthesized using MPTES (5g, 0.017mol) as described above. The zirconium was stabilized using 3 equivalents of methacrylic acid compared to the mole of zirconium propoxide. For example, to synthesize PSQ with 5 mol% zirconium, the zirconium propoxide (70 wt% in propanol, 0.424g, 0.0009mol) was mixed with methacrylic acid (0.235g, 0.0027mol) in a sealed glass vial. This slightly yellow solution was stirred at room temperature for 24 hours. After that, it was mixed with the prepared PSQ to form a homogenous solution. For samples containing 5 mol% and 7.2 mol% zirconium, all the volatiles were removed under vacuum to give a yellow-orange viscous oil. For sample containing 10 mol% zirconium, 0.15 g of dodecanol was added as a viscosity controller before removing all volatiles. The resulting resin is yellow-orange viscous oil.

**Prepare thin film for refractive index measurements**

The refractive index of different PSQs were measured using an ellipsometer. To measure the refractive index, thin films were prepared as the procedure described below.

The as-synthesized oil material was diluted using THF with the volume ratio of 1:2. Quartz thin slides were used as the substrate since it has good stability within the temperature range during thermal treatment. The diluted solution was first dropped and spread on the whole surface of the quarts slide with proper dimension. The spin coating was done with 4000 rpm for 1min. The obtained coated slides were then exposed under UV for 10min to cure the thin film. The thin films prepared using this procedure have thicknesses between 1500 to 1700 nm. After the curing, the quartz with thin film was heated to 650 ℃ with a rate of 1 ℃/min and kept at 650 ℃ for 2h to get inorganic glass thin films. No crack was observed for all the glass films prepared using this procedure. After the thermal treatment, the films have thicknesses between 400 to 600 nm.

**Two-photon printing of PSQ**

The printing system contains a 780 nm femtosecond fiber laser with 150 fs pulse, 77MHz, and a maximum power of 130 mW. The full-width half maximum (FHWM) of the beam was 10 mm, 90% filling the objective (NA=1.3). The component was printed with a 1.17 nJ pulse energy and 16900 μm/s on the quartz substrate.

After printing, the uncured resin was washed using PGMEA and ethanol and dried before thermal treatment.

**Thermally convert printed parts to inorganic glass**

The printed parts were heated in a furnace to 650 ℃ with a heating rate of 1 ℃/min in air. The sample was then kept at 650 ℃ for 2h and gradually cooled down to room temperature.

**Printing optics with multi-components through self-welding**

Individual single lens optics were printed using different PSQ resin (PSQ-MPTES, PSQ-5%Zr). After printing and washing, the PSQ-5%Zr sample was carefully transferred on top of the PSQ-MPTES sample using a micro fiber. The stacked micro-optics were placed into furnace to go thorugh pyrolysis process described above. The welding process finished automatically during the pyrolysis process.

## Author contributions

P.Y. and Z.H. contributed equally to this work. P.Y., Z.H., D.A.L., and R.L. conceived the idea and designed the study. P.Y. and D.A.L. designed the material. P.Y. prepared, and characterized the materials. Z.H. designed the optics, performed printing experiments, and conducted the imaging experiments. P.Y., Z.H., D.A.L., and R.L. analyzed and interpreted the result, and wrote the manuscript.

## Acknowledgment

This work was supported by National Cancer Institute R21CA268190.

# Solvent-Free Silsesquioxane Self-Welding for 3D Printing Multi-Refractive Index Glass Objects

Piaoran Ye,[1]† Zhihan Hong,[1*]† Douglas A. Loy,[2,3] Rongguang Liang[1*]

[1]Wyant College of Optical Sciences, The University of Arizona, 1630 E. University Blvd, Tucson, Arizona 85721, USA

[2]Department of Chemistry&Biochemistry, The University of Arizona, 1306 E. University Blvd, Tucson, Arizona 85721-0041, USA

[3]Department of Materials Science&Engineering, The University of Arizona, 1235 E. James E. Rogers Way, Tucson, Arizona 85721-0012, USA

*Corresponding author: Zhihan Hong (zhihanhong@arizona.edu)

Rongguang Liang (rliang@optics.arizona.edu)

† These two authors contributed equally to this work.

**Characterization**

Infrared (IR) spectra were obtained with a Thermo Scientific Nicolet iS50R using a Harrick MVP-Pro™ Single Reflection ATR Microsampler. 1D and 2D nuclear magnetic resonance (NMR) was obtained using a Bruker NEO 500 MHz automatic system. Scanning Electron Microscope (SEM) images were taken using FEI Inspect Scanning Electron Microscope. Raman spectra was obtained through a Renishaw InVia system using 785nm laser. The surface profile was measured by Zygo Newview 8300 white light interference microscope.

The viscosity of all samples were measured using RheoSense Small Sample m-VROC equiped with C seiral chip. The max reliable viscosity this instrument can measure is 14000 m-Pas

The refractive index of thin film was measured using Woollam m2000 Spectroscopic Ellipsometer. The measurements were conducted with 50°, 55°, and 60°. The refractive index (n) and extinction coefficient (k) were calculated by the software through Cauchy model.

**Discussion**

Preparation of Zr-MAA complex

We mixed zirconium propoxide (70 wt% in 1-propanol) with 3 equivalents of MAA to form the Zr-MAA complex and used $^1$H NMR and $^{13}$C NMR to characterize the product. No recrystallization or other purification was attempted. From previous literature, it is likely that the final form of the complex would be $Zr_x(OCH_2CH_2CH_3)_y(OOC_4H_5)_z$, where x doesn't equal to 1. The broad peaks in both $^1$H NMR (6.18 ppm, 5.49 ppm, 1.85 ppm) and $^{13}$C NMR (137.7 ppm, 126.4 ppm, 17.6 ppm) indicate that almost all the MAA have bonded to the Zr. No peak belonging to pristine MAA molecules was observed (Fig. S8 & S9). One byproduct, propyl methacrylate, was found in the final product mixture. This ester was formed from the esterification between MAA and propanol. This can be confirmed from HSQC, HMBC, and COSY 2D spectra (Fig. S10, S11, and S12). Meanwhile, this phenomenon was also reported in a previous literature and it is difficult to avoid.[1] Applying vacuum to the Zr-MAA complex mixture removed most of the propanol to result a solid product which can be re-dissolved in organic solvents. Figure S13b shows that after vacuuming, the sharp peaks (1.59 ppm, 0.91 ppm) and the broad OH peak (2.56 ppm) belonging to propanol disappeared. Meanwhile, most of the propyl methacrylate remained (Fig. S13a) due to the relatively high boiling point. However, based on the integration from the $^1$H NMR, the total ratio of propyl methacrylate is less than 10 mol% compared to the number of bonded MAA molecules.

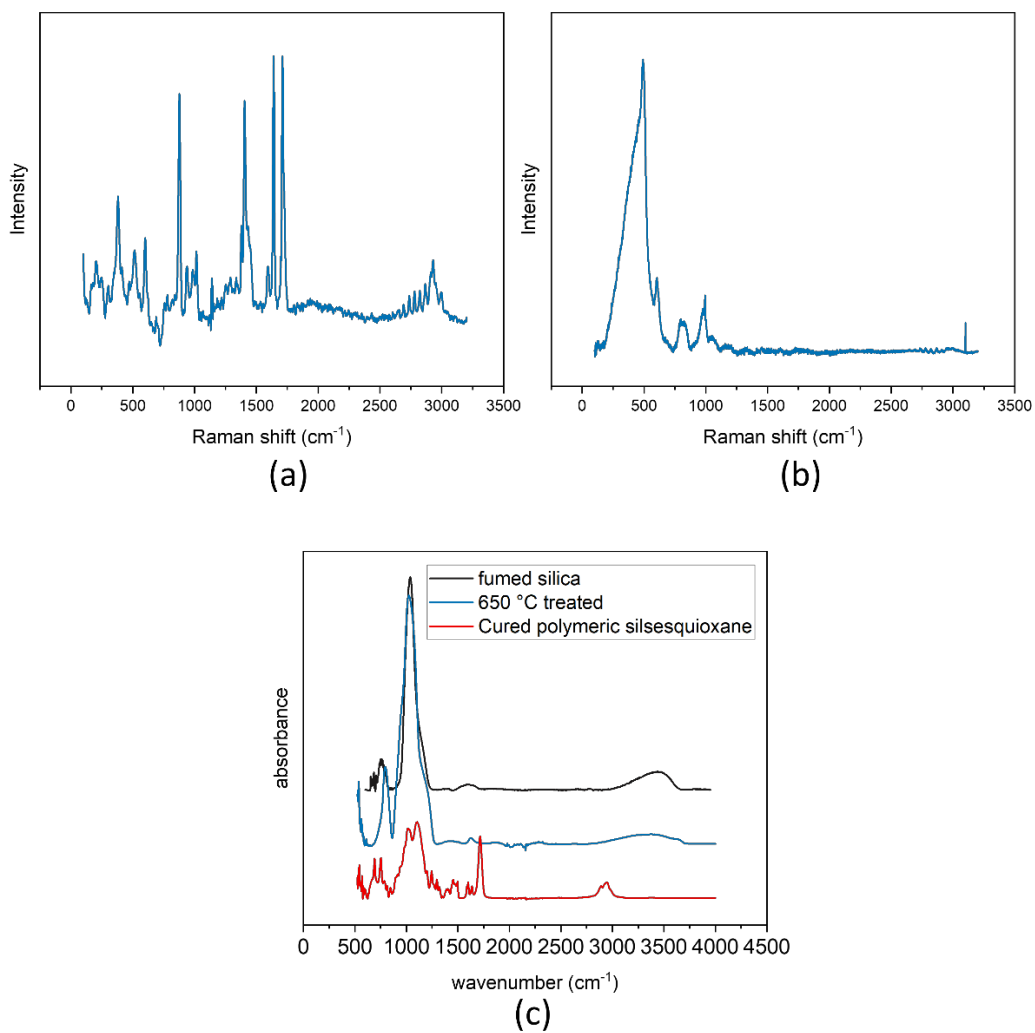

Figure S1. (a) Raman spectrum of as-printed PSQ. (b) Raman spectrum of thermally treated PSQ. (c) Comparison FTIR spectra of fumed silica powder, thermally treated PSQ, and photo-cured PSQ.

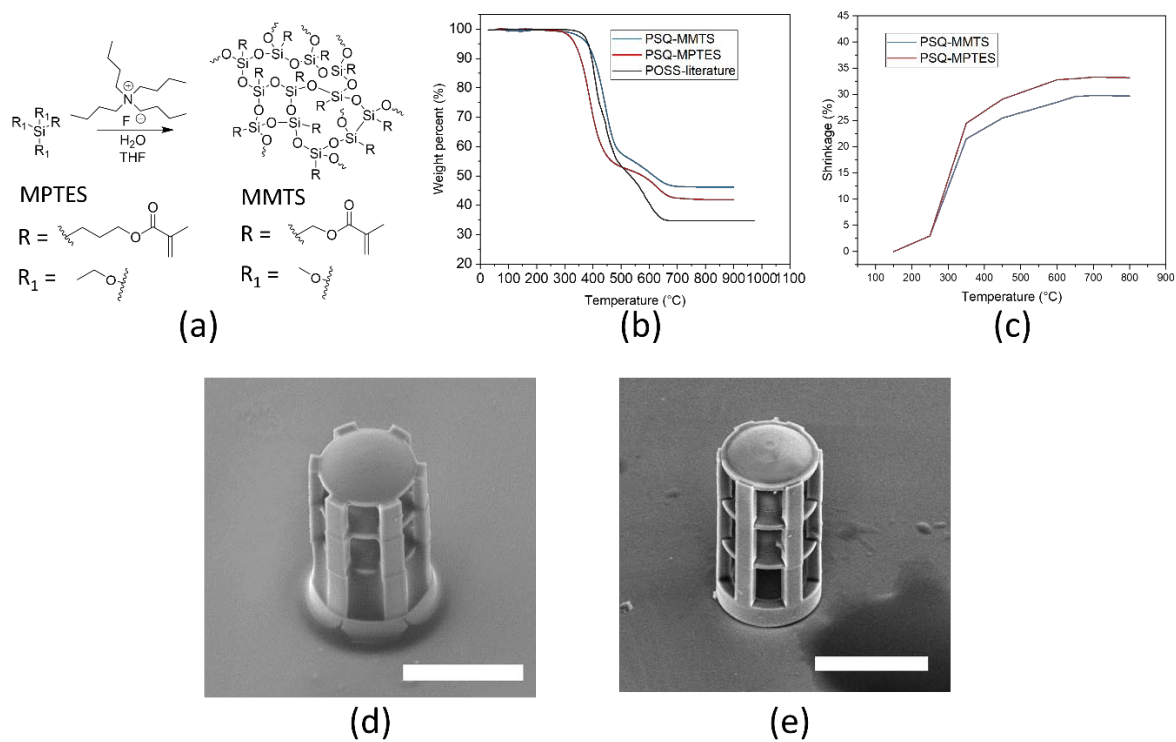

Figure S2. (a) Scheme of using different monomers (MPTES or MMTS) to synthesize PSQ. (b) TGA results of UV-cured PSQs. The curve of POSS was drawn using the data from literature.(*20*) (c) The measured shrinkage at different temperatures during the thermal treatment. (d) SEM images of glass triple lens micro-objective printed using PSQ-MPTES and (e) PSQ-MMTS, respectively. These two objectives have slightly different structure with different diamensions. Scale bar: 200 μm.

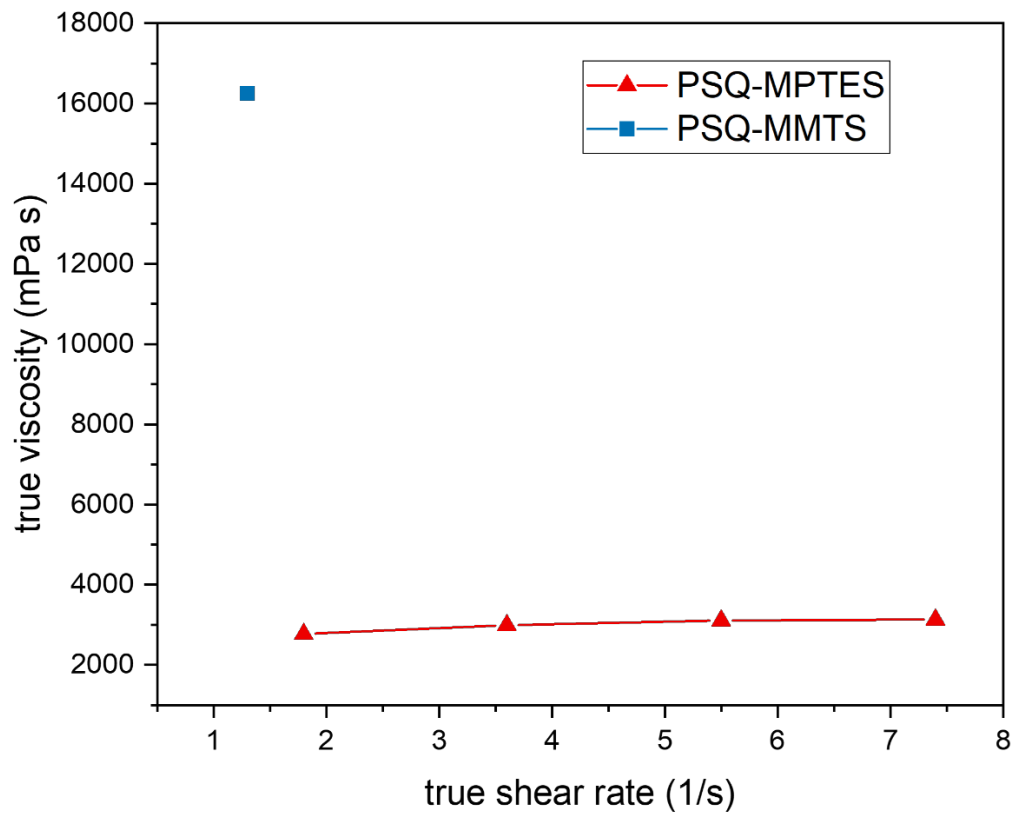

Figure S3. The viscosity of PSO-MPTES and PSQ-MMTS. The viscosity of PSQ-MMTS was measured only under one shear rate since the viscosity at higher shear rate is higher than the instrument's safety limitation. It is noteworthy that 16000 mPa-s is a viscosity that is higher than the instrument's reliable measurement range.

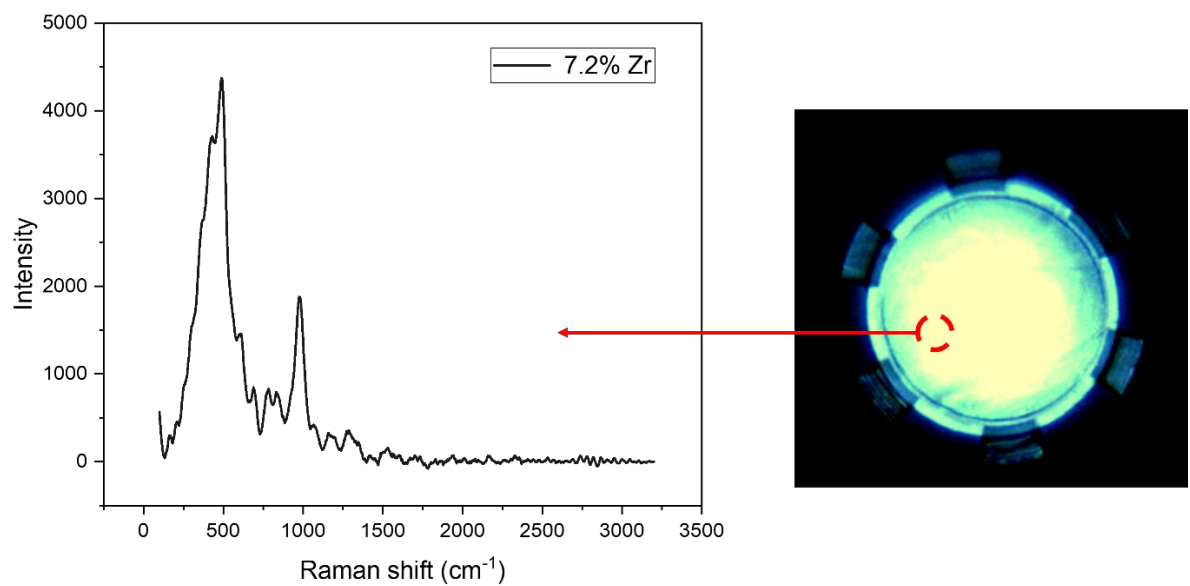

Figure S4. Micro-Raman spectrum of PSQ-MPTES containing 7.2% Zr after thermal treatment. The spectrum was taken wihin the red circle region.

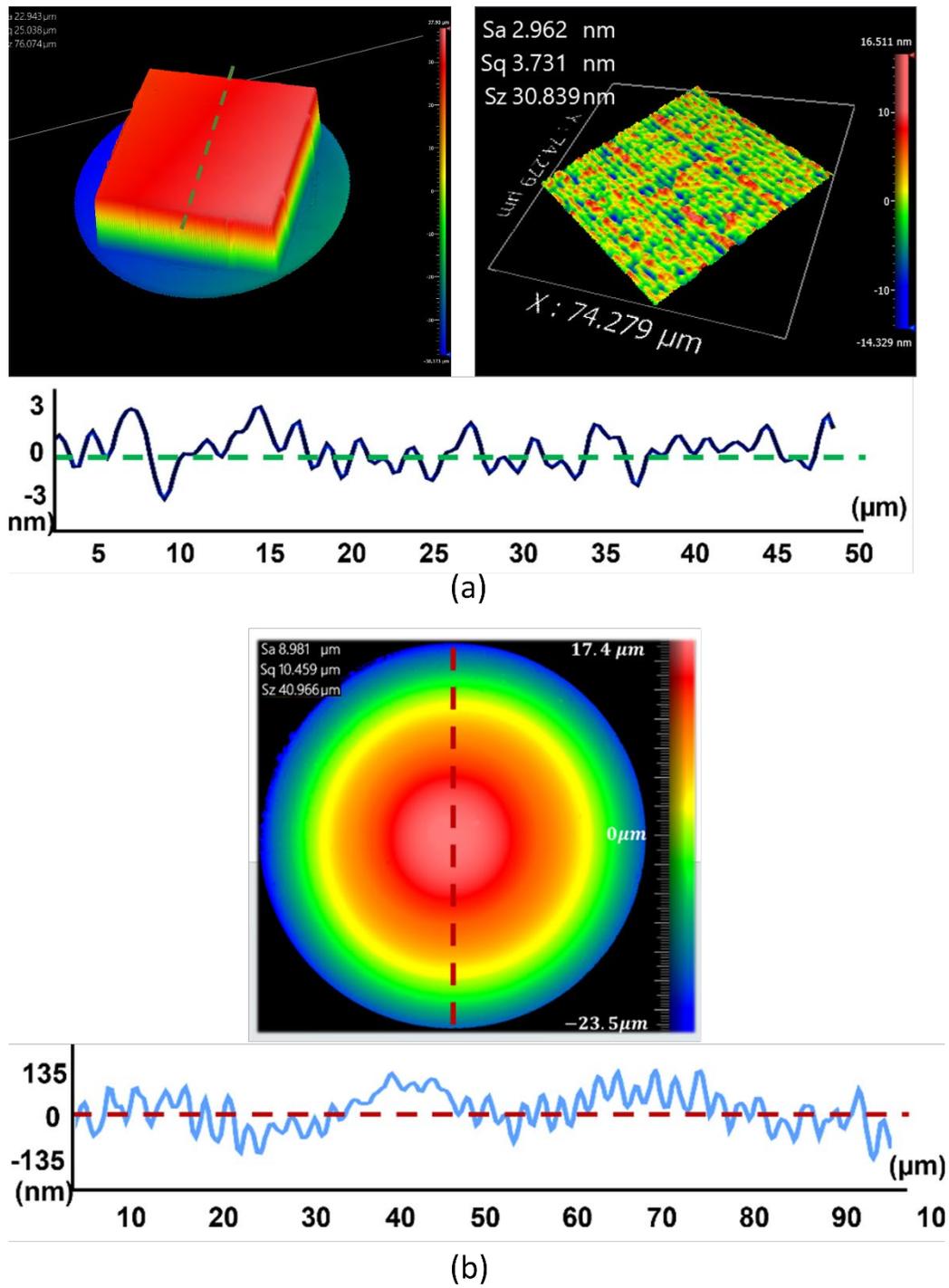

Figure S5. (a) Surface measurements of a 2PP printed flat element. (b) Surface measurements of a 2PP printed aspherical structure.

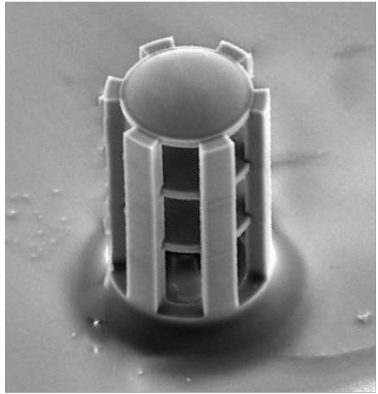 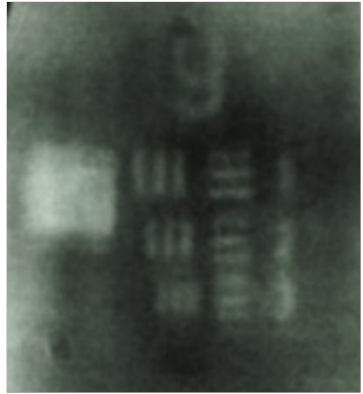 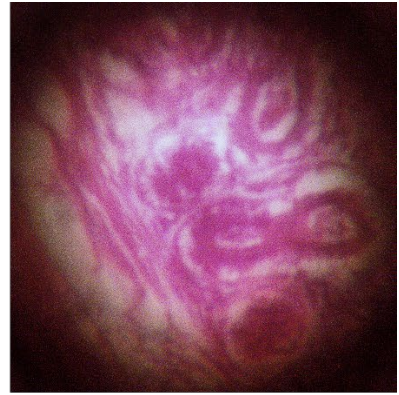

(a) (b) (c)

Figure S6. (a) The SEM image of the glass micro-objective containing 7.2% Zr. (b) The image of 1951 USAF target (Group 9). (c) The image of Human resting mammary gland captured by optical system fabricated using micro-objective shown in (a)

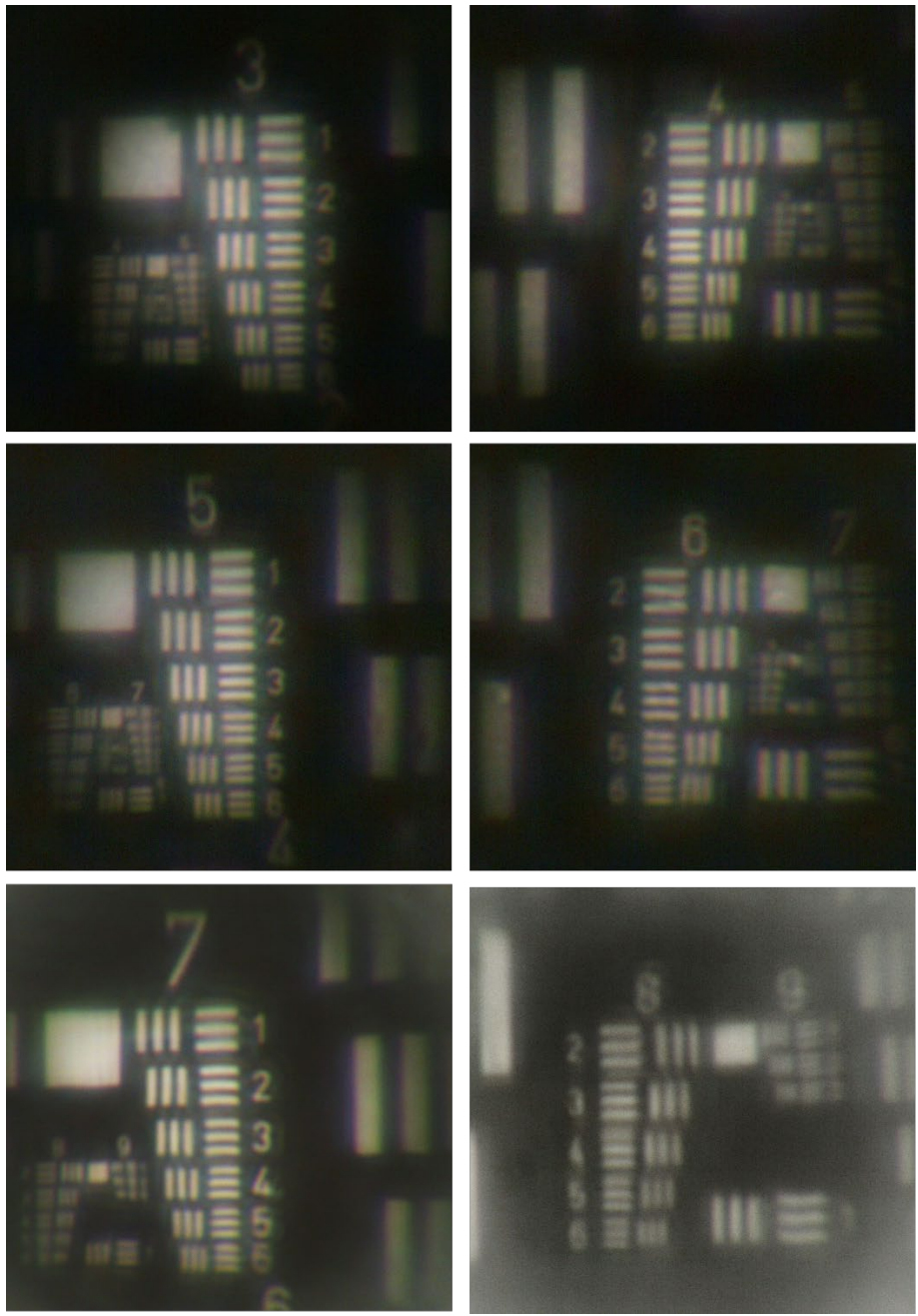

Figure S7. 1951 USAF target imaging performance of micro-objective shown in Figure S6.

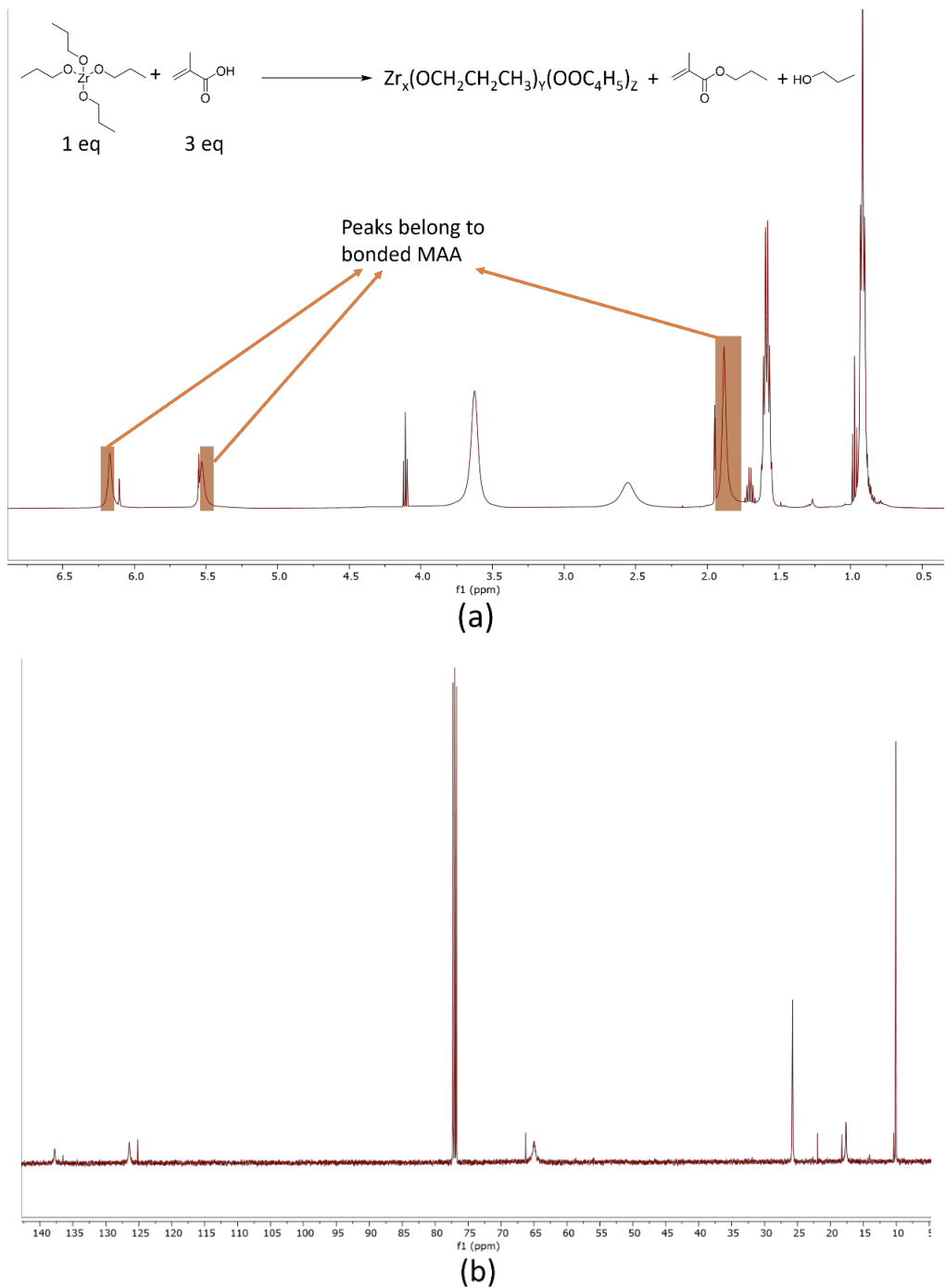

Figure S8. $^1$H and $^{13}$C NMR of the mixture formed by 1 equivalent of zirconium propoxide and 3 eq of MAA (24h mixing).

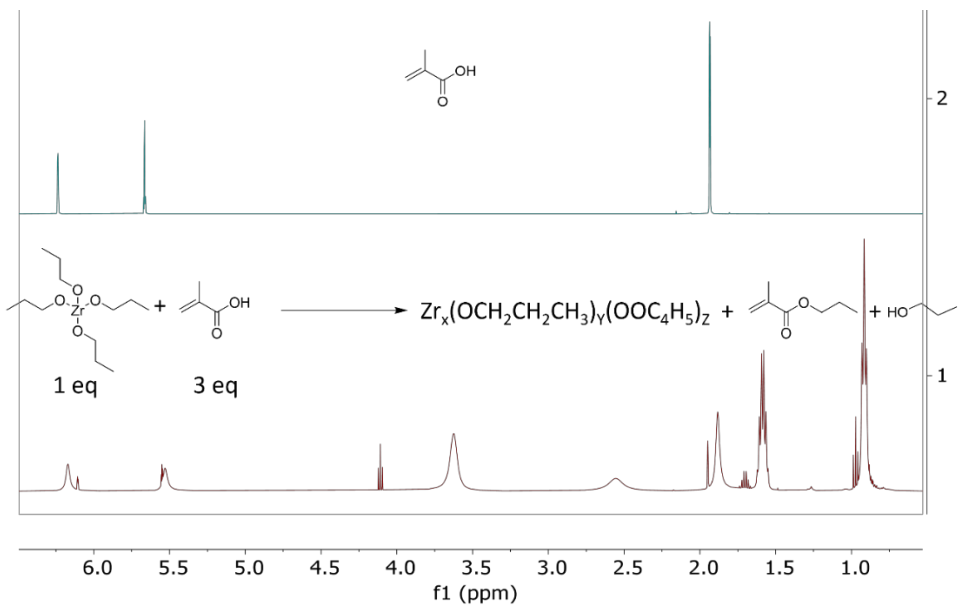

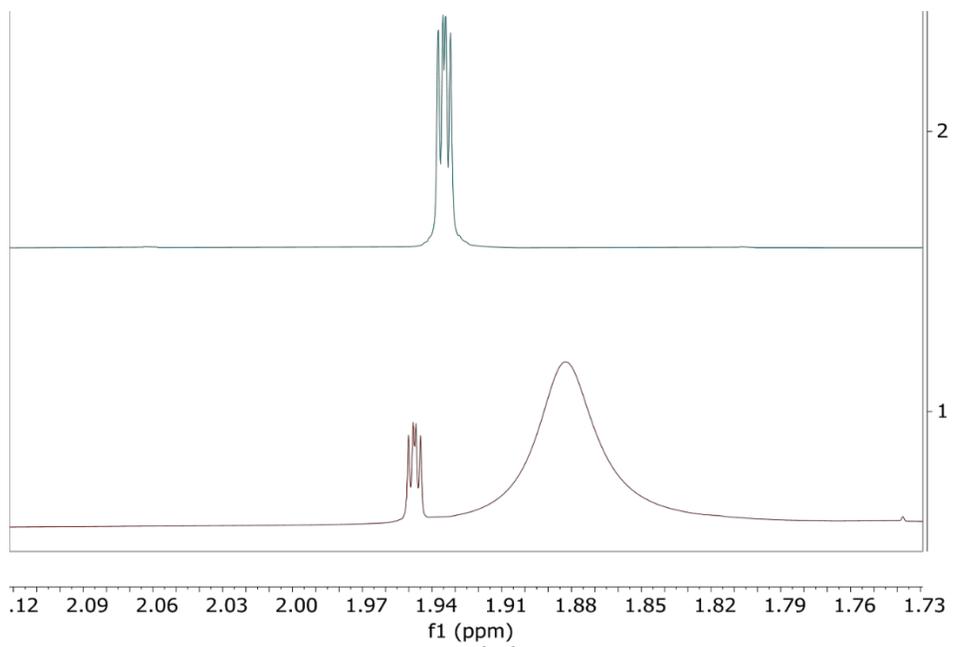

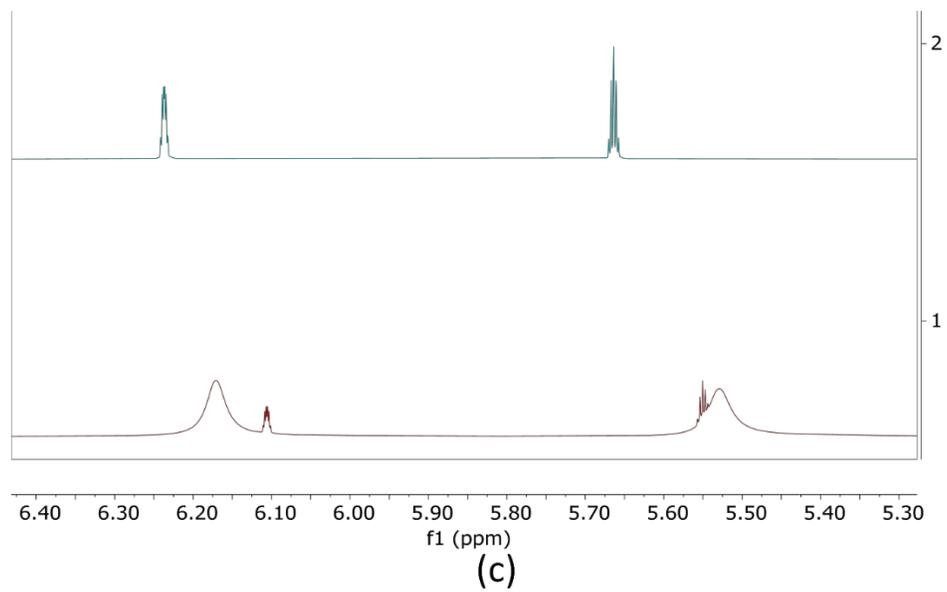

Figure S9. NMR spectra comparison between the zirconium propoxide-MAA mixture and the pristine MAA.

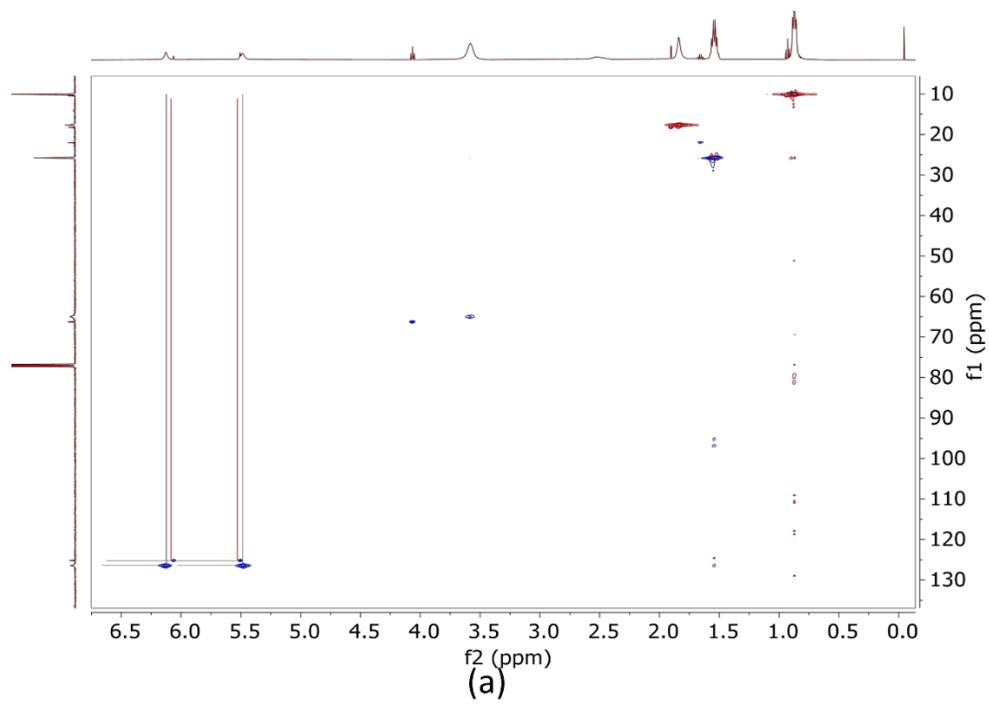

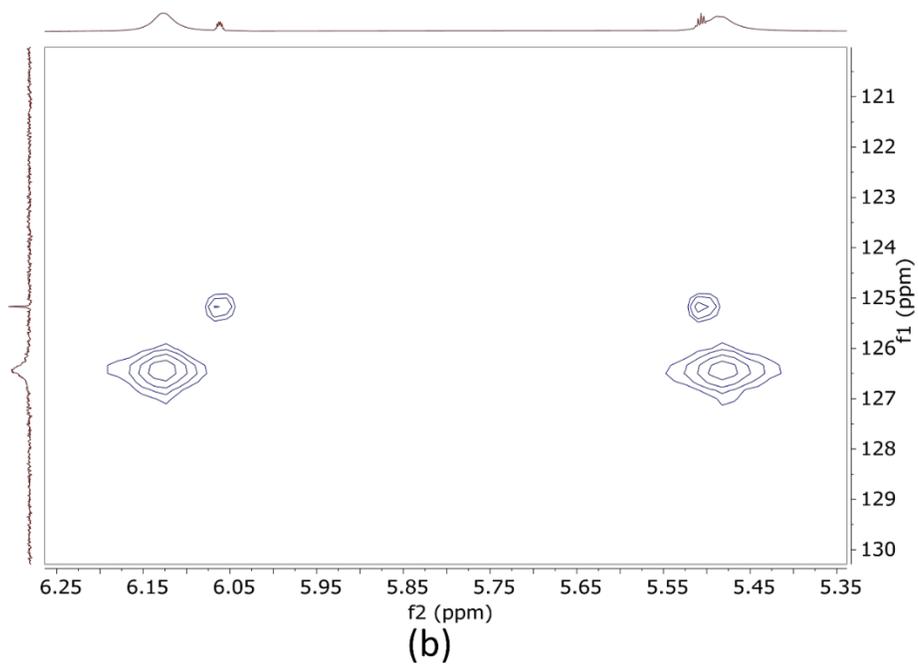

Figure S10. HSQC spectrum of zirconium propoxide-MAA mixture

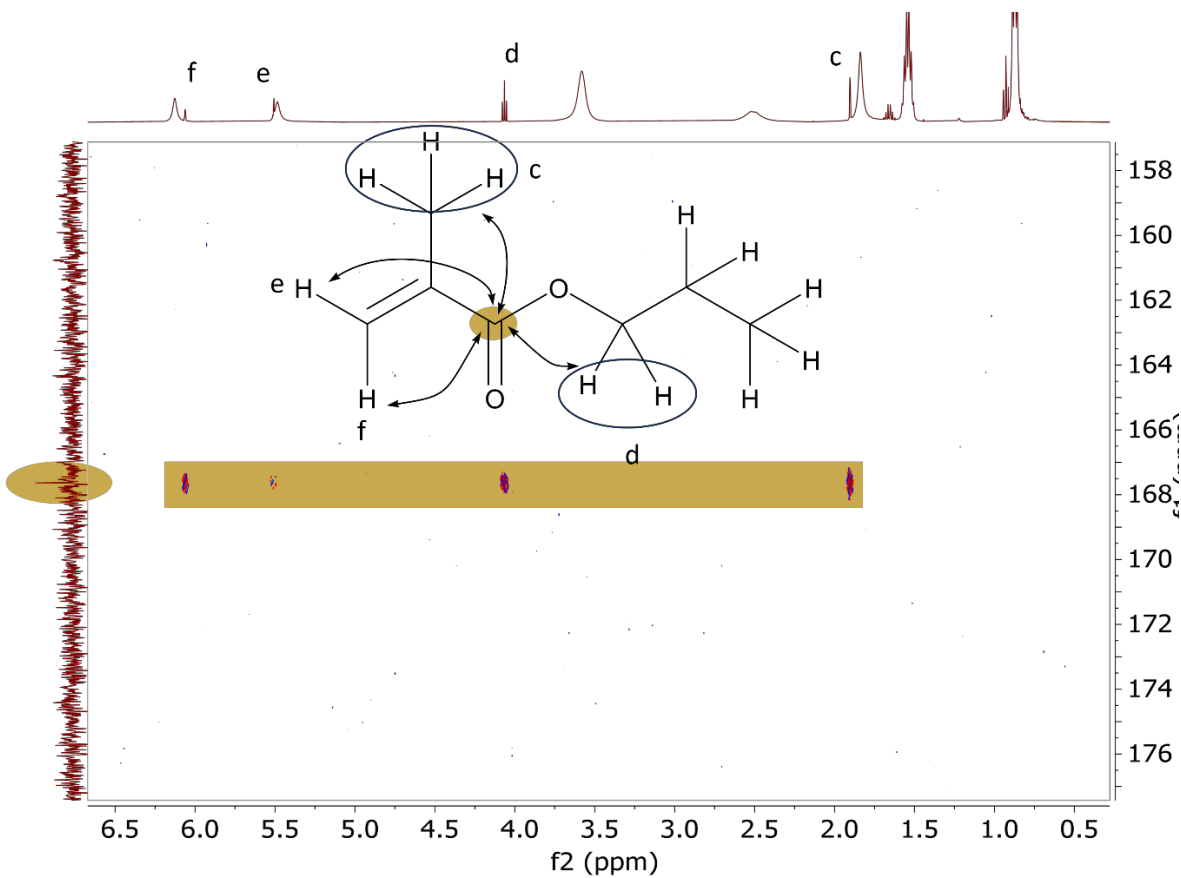

Figure S11. HMBC spectrum of zirconium propoxide-MAA mixture.

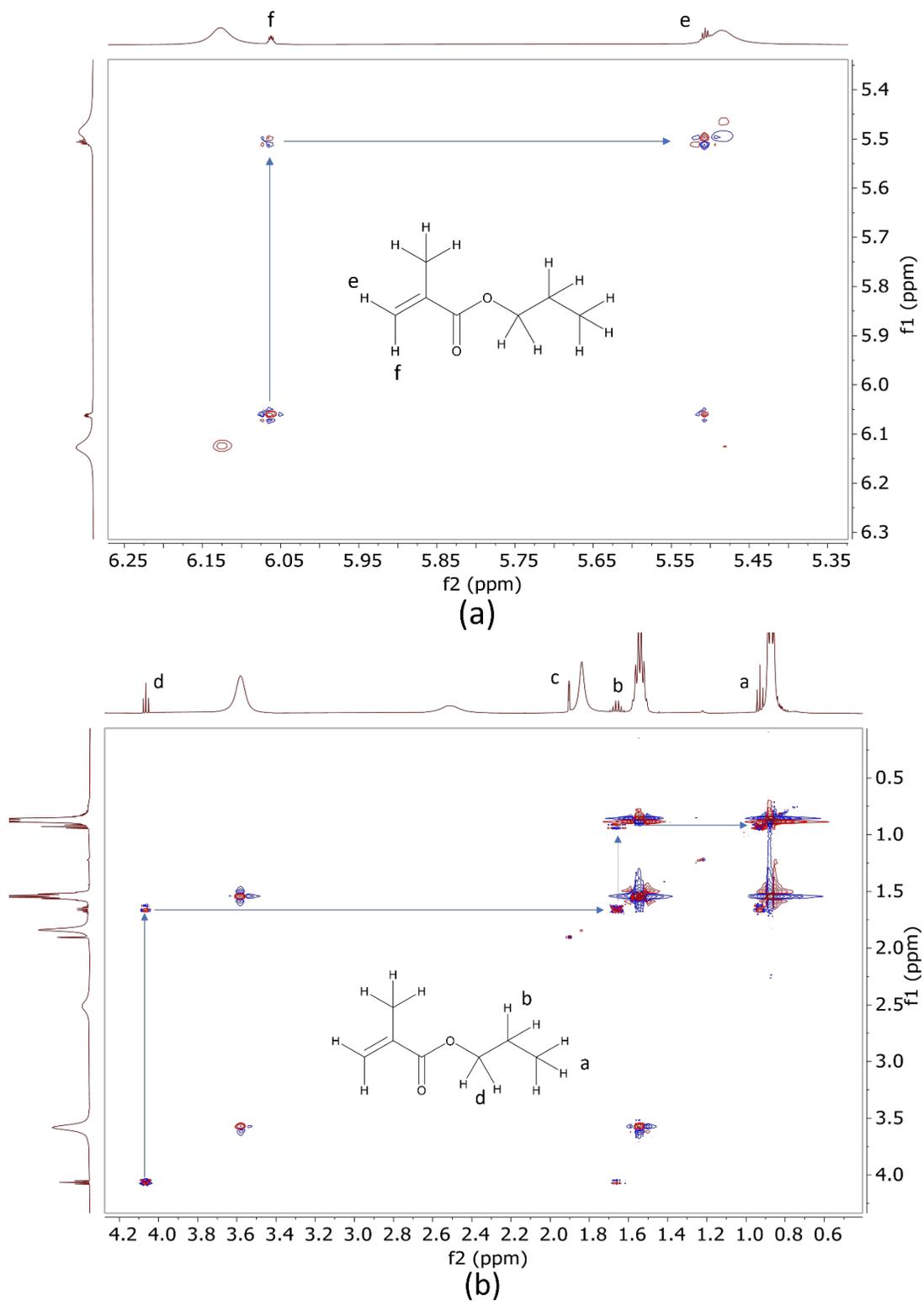

Figure S12. COSY spectrum of zirconium propoxide-MAA mixture.

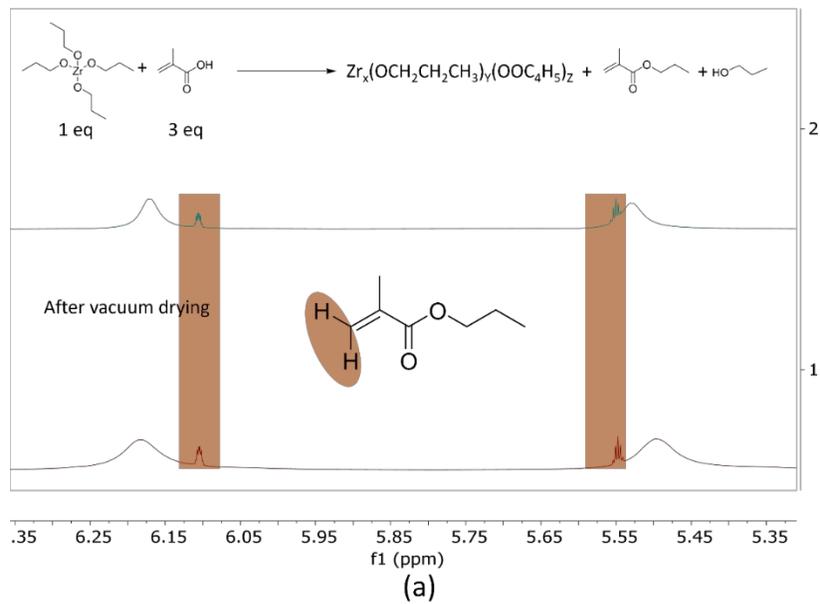

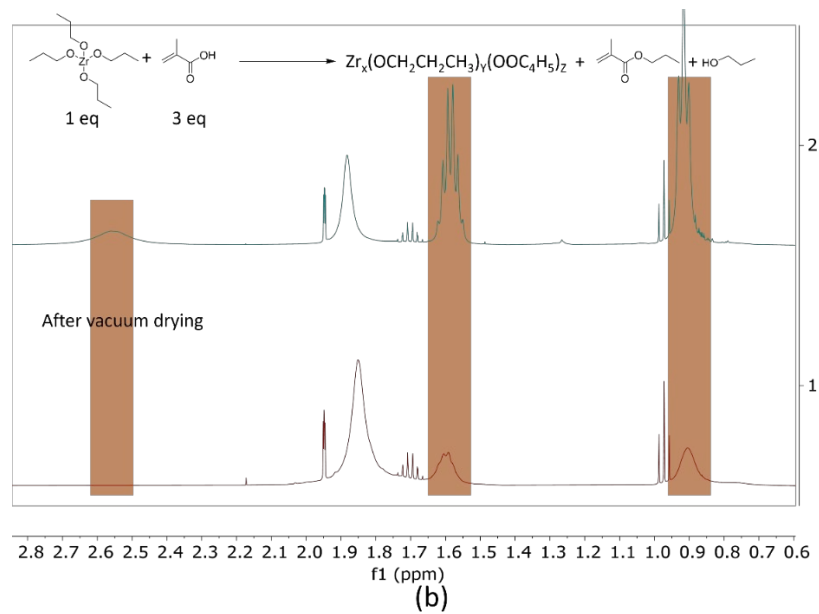

Figure S13. Comparison ¹H NMR spectra of zirconium propoxide-MAA mixture before and after vacuum drying.